\documentclass[11pt]{article}
\usepackage{xspace}
\usepackage{graphicx}
\usepackage{amsmath}
\usepackage{amssymb}
\usepackage{color}

\definecolor{Red}{rgb}{1,0,0}
\definecolor{Green}{rgb}{0,1,0}
\definecolor{Blue}{rgb}{0,0,1}
\definecolor{Black}{rgb}{0,0,0}

\def\beq{\begin{equation}}
\def\eeq#1{\label{#1}\end{equation}}
\def\eeqn{\end{equation}}

\def\beqa{\begin{eqnarray}}
\def\eeqa#1{\label{#1}\end{eqnarray}}
\def\eeqan{\end{eqnarray}}

\let\bar=\overbar

\def\Dslash{\not{\hbox{\kern-4pt $D$}}}
\def\dslash{\not{\hbox{\kern-2pt $\del$}}}

\def\msb{{\bar{\ssstyle M \kern -1pt S}}}

\newcommand\numu     {\ensuremath{\nu_{\mu}}\xspace}

\newcommand\mrad   {\,mrad\xspace}
\newcommand\m      {\,m\xspace}
\newcommand\km     {\,km\xspace}
\newcommand\ton    {\,ton\xspace}
\newcommand\kton   {\,kton\xspace}

\newcommand\ns     {\,ns\xspace}
\newcommand\degree {\ensuremath{^{\circ}}\xspace}

\newsavebox\IBoxA \newsavebox\IBoxB \newsavebox\IBoxC \newlength\IHeightA
\newcommand\ThreeFig[9]{% Image1 Caption1 Label1 Image2 ...
  \sbox\IBoxA{\includegraphics[width=0.32\columnwidth]{#1}}
  \sbox\IBoxB{\includegraphics[width=0.32\columnwidth]{#4}}%
  \sbox\IBoxC{\includegraphics[width=0.32\columnwidth]{#7}}%
  \ifdim\ht\IBoxA>\ht\IBoxB
    \setlength\IHeightA{\ht\IBoxB}\else\setlength\IHeightA{\ht\IBoxA}\fi%
  \ifdim\ht\IBoxC<\IHeightA
    \setlength\IHeightA{\ht\IBoxC}\fi%
  \begin{figure}[!ht]
  \begin{center}
  \minipage[t]{0.32\columnwidth}\centering
  \includegraphics[height=\IHeightA]{#1}
  \caption{#2}\label{#3}
  \endminipage\hfill
  \minipage[t]{0.32\columnwidth}\centering
  \includegraphics[height=\IHeightA]{#4}
  \caption{#5}\label{#6}
  \endminipage\hfill
  \minipage[t]{0.32\columnwidth}\centering
  \includegraphics[height=\IHeightA]{#7}
  \caption{#8}\label{#9}
  \endminipage 
  \end{center}
  \end{figure}%
}

\newsavebox\IBoxD \newsavebox\IBoxE \newlength\IHeightB
\newcommand\TwoFig[6]{% Image1 Caption1 Label1 Image2 ...
  \sbox\IBoxD{\includegraphics[width=0.45\columnwidth]{#1}}
  \sbox\IBoxE{\includegraphics[width=0.45\columnwidth]{#4}}%
  \ifdim\ht\IBoxD>\ht\IBoxE
    \setlength\IHeightB{\ht\IBoxE}\else\setlength\IHeightB{\ht\IBoxD}\fi%
  \begin{figure}[!ht]
  \begin{center}
  \minipage[t]{0.45\columnwidth}\centering
  \includegraphics[height=\IHeightB]{#1}
  \caption{#2}\label{#3}
  \endminipage\hfill
  \minipage[t]{0.45\columnwidth}\centering
  \includegraphics[height=\IHeightB]{#4}
  \caption{#5}\label{#6}
  \endminipage\hfill
  \end{center}
  \end{figure}%
}

\def\Title#1{\begin{center} {\Large {\bf #1} } \end{center}}

\begin{document}

\Title{Construction of the CHIPS-M prototype and simulations of a 10 kiloton module}

\bigskip\bigskip

%+\addtocontents{toc}{{\it D. Reggiano}}
%+\label{ReggianoStart}

\begin{raggedright}  

%% Authors - you should specify at least one author as follows.
{\it Andrew Perch\index{Perch, A.},\\
Department of Physics and Astronomy\\
University College London\\
Gower Street\\
London\\
WC1E 6BT\\}
%% In case you want to have more than one author please follow the format
%% shown below, listing the individual authors AND also making sure
%% that each author is given a unique index entry.
%Someone Else\index{Else, S.}, {\it Another University}\\

\end{raggedright}
\vspace{1.cm}

{\small \begin{flushleft} \emph{To appear in the proceedings of the Prospects in
Neutrino Physics Conference, 15 -- 17 December, 2014, held at Queen Mary
University of London, UK.} \end{flushleft} }

\section{Introduction}
CHIPS is an R\&D experiment intending to produce large, low-cost water Cherenkov
detectors that can be used to study long-baseline neutrino oscillations.  Such detectors
are traditionally constructed underground, incurring large civil engineering
costs in order to excavate or renovate a suitable cavern.  CHIPS instead
proposes to construct \textbf{CH}erenkov detectors \textbf{i}n Mine
\textbf{P}it\textbf{s}, by submerging a detector at the bottom of deep bodies of
water on the surface of the Earth. 

In the summer of 2014, a small 26\ton prototype detector was deployed in a
former iron quarry in Northern Minnesota, 7\mrad off-axis from the NuMI beam.
This poster describes the construction and deployment of this prototype, and
simulations of a larger 10\kton module.

\section{CHIPS experiment} 
As well as reducing engineering costs, constructing the detector in a lake will
provide the detector with sufficient overburden.  For a 40\kton detector at a
depth of 40\m the cosmic ray event rate is estimated to lead to a mean dead time
per NuMI beam spill of 250\ns (or 2.5\%)\cite{Adamson:2013xka}.

The detector is filled with filtered lake water, enclosed in a
waterproof, opaque plastic geomembrane.  The design attemps to use commercially
available materials where possible: studies for the full-size modules are
currently investigating the use of a ``space frame'' of interlocking
tubes, and fibreglass panels used by radar dome manufacturers.

The experiment is located at the Wentworth 2W pit, near to Hoyte Lakes,
Minnesota.  A former iron quarry, the pit is 60\m deep and situated 7\mrad
off-axis from Fermilab's NuMI neutrino beam at a baseline of 712\km.  This
exposes CHIPS to a lower peak beam energy than MINOS (on-axis) but a higher
flux compared to NOvA (14\mrad).

\section{CHIPS-M}
A 26\ton prototype detector, CHIPS-M (``Model''), was constructed during the
summer of 2014 by a team largely made up of students and postdocs.  CHIPS-M is an octagonal detector, 3.5\m tall by 3.2\m in
diameter, comprising a rigid frame of aluminium stage truss and stainless
steel unistrut surrounded by an opaque plastic geomembrane.  The membrane is
white on the outside and black on the inside (allowing for a possible veto region
outside the main volume) and is tightly clamped between aluminium
batten bars and the main frame of the detector with waterproof tape in between.

The detector is instrumented with five DOMs: Digital Optical
Modules\cite{Abbasi:2008aa} consisting of a pressure sphere containing a 10"
photomultiplier tube and electronics for high voltage supply, data acquisition,
etc., borrowed from the IceCube experiment.  These are attached to the frame
using a pivoting mount that has floats attached to point the DOMs' photocathodes
at a 45\degree angle to the beam.  IceCube's DOMs are designed to point
downards, so this angle was chosen to reduce the risk of the optical gel
shearing compared to facing the beam directly. GPS timing is used so that
timestamps from the DOMs can be compared to NuMI beam spills.

Two pressure cylinders containing cameras and environmental monitoring sensors
are also present, one inside the detector and one outside.  These proved
invaluable in the deployment and filling of the detector.

\ThreeFig         {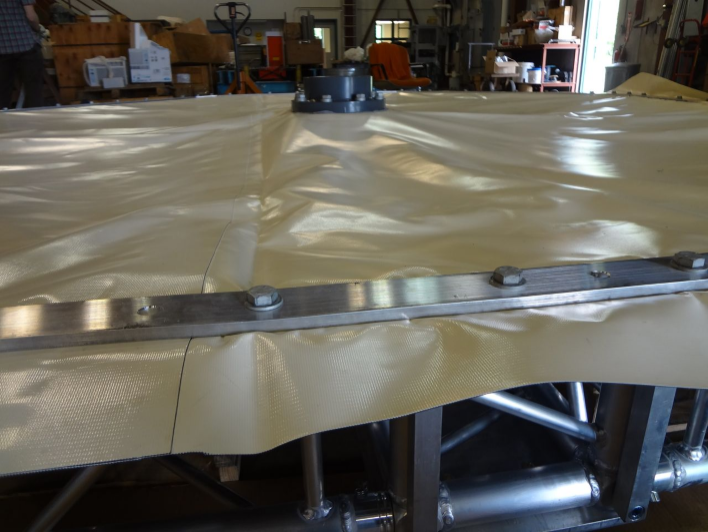} {Aluminium batten bars which clamp the liner and
         waterproof tape to the frame} {fig:battenbar}
         {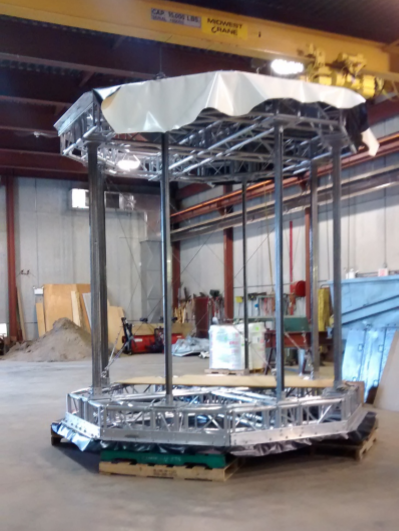}     {Top and bottom stage truss sections covered with
liner and held apart by stainless steel unistrut} {fig:frame}
{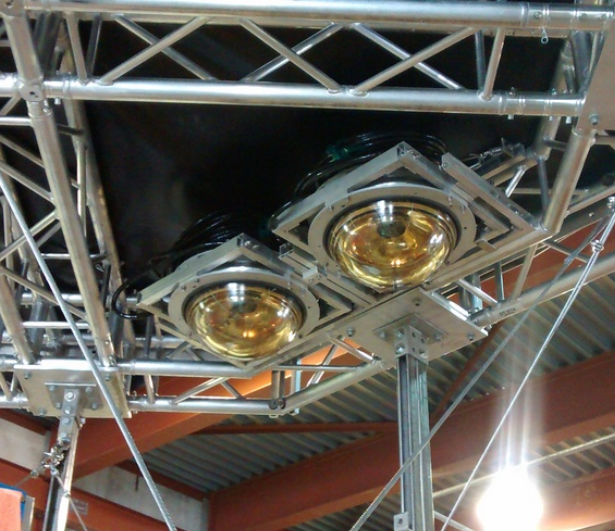}      {Two DOMs mounted at the top of the detector} {fig:doms}

The prototype is intended to test the suitability of the materials chosen, in
particular their durability underwater and the performance of the liner and
sealing tape.  It will also provide indications as to how sophisticated a water
filtration system is required in order to achieve acceptable water clarity in a
10\kton module.  The detector will be retrieved and assessed in June 2015,
fitted with additional PMTs (a flat panel of 3" tubes derived from KM3NeT's
31-PMT optical module) and environmental sensors, and resubmerged for a second
year of tests.

\section{Data from CHIPS-M}
CHIPS-M is submerged and has been taking data since the beginning of August,
observing Cherenkov light primarily from cosmic ray muons (beam events are
difficult to distinguish in a detector of this size).  For these studies,
an event is defined as a coincidence between the four DOMs at the top of the
detector, with each hit above 5 photoelectrons and within 15ns of the first.
The rate of these events is shown in Figure \ref{fig:eventrate}.

\TwoFig{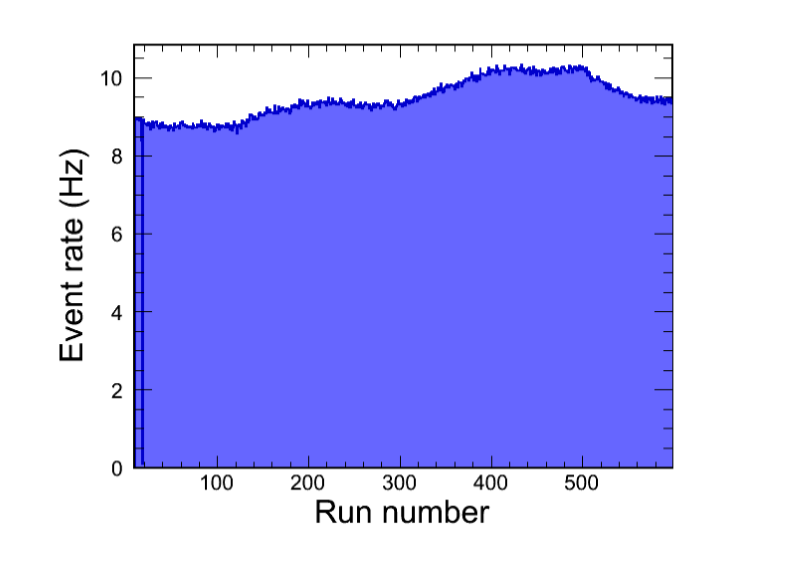}{Rate of four DOM coincidences observed at
CHIPS-M}{fig:eventrate}
       {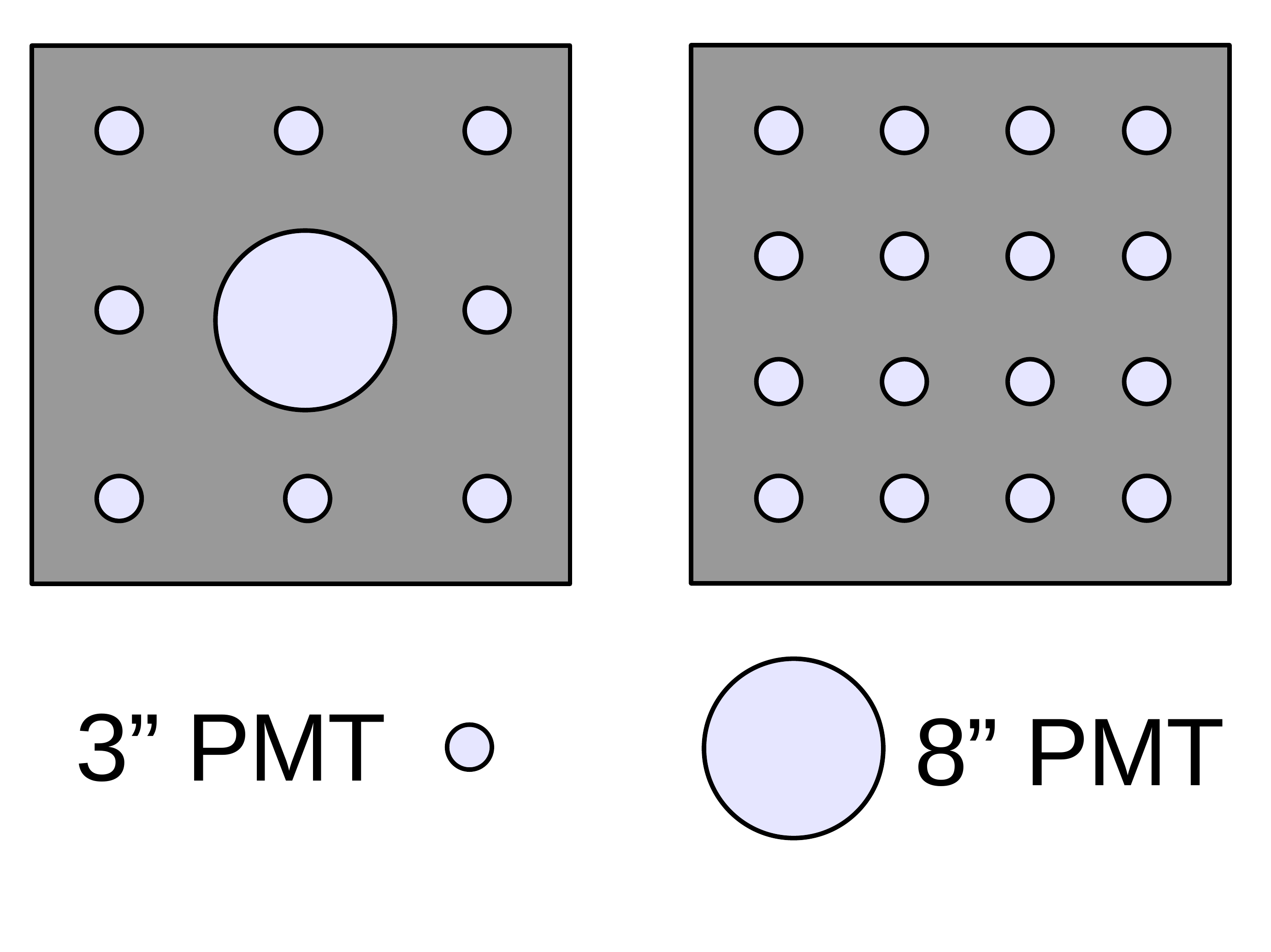}{Schematic of two patterns that can be
       compared using the simulation}{fig:layout_schematic}
       
The DOMs also contain an LED flasher board.  Studies into using this board to
measure the water clarity over time are ongoing.

\section{CHIPS-10}
The intention of CHIPS is to design a 10\kton module for which construction
could begin in 2016.  The baseline design is a cylinder 25\m in diameter and
20\m high, constructed from tiling modules of fibreglass, or a spaceframe with
panels of geomembrane liner. This would be constructed in layers before finally sealing together the walls
and the top and sinking it to the bottom of the pit.  

An advantage of such a design is that the full-size detector need not be constructed at once,
but can be retrieved to add additional layers as funding and seasonal
ice coverage allow.  This also permits the detector to be redeployed in the
future LBNF beam.

The 10\kton detector would have a cosmic ray veto region of outward
facing PMTs, and be able to accomodate different types of PMT to operate as a
test-bed for optical modules exposed to a real neutrino beam.

\section{Simulations of CHIPS-10}
\begin{figure}[!ht]
\begin{center}
\includegraphics[width=0.75\columnwidth]{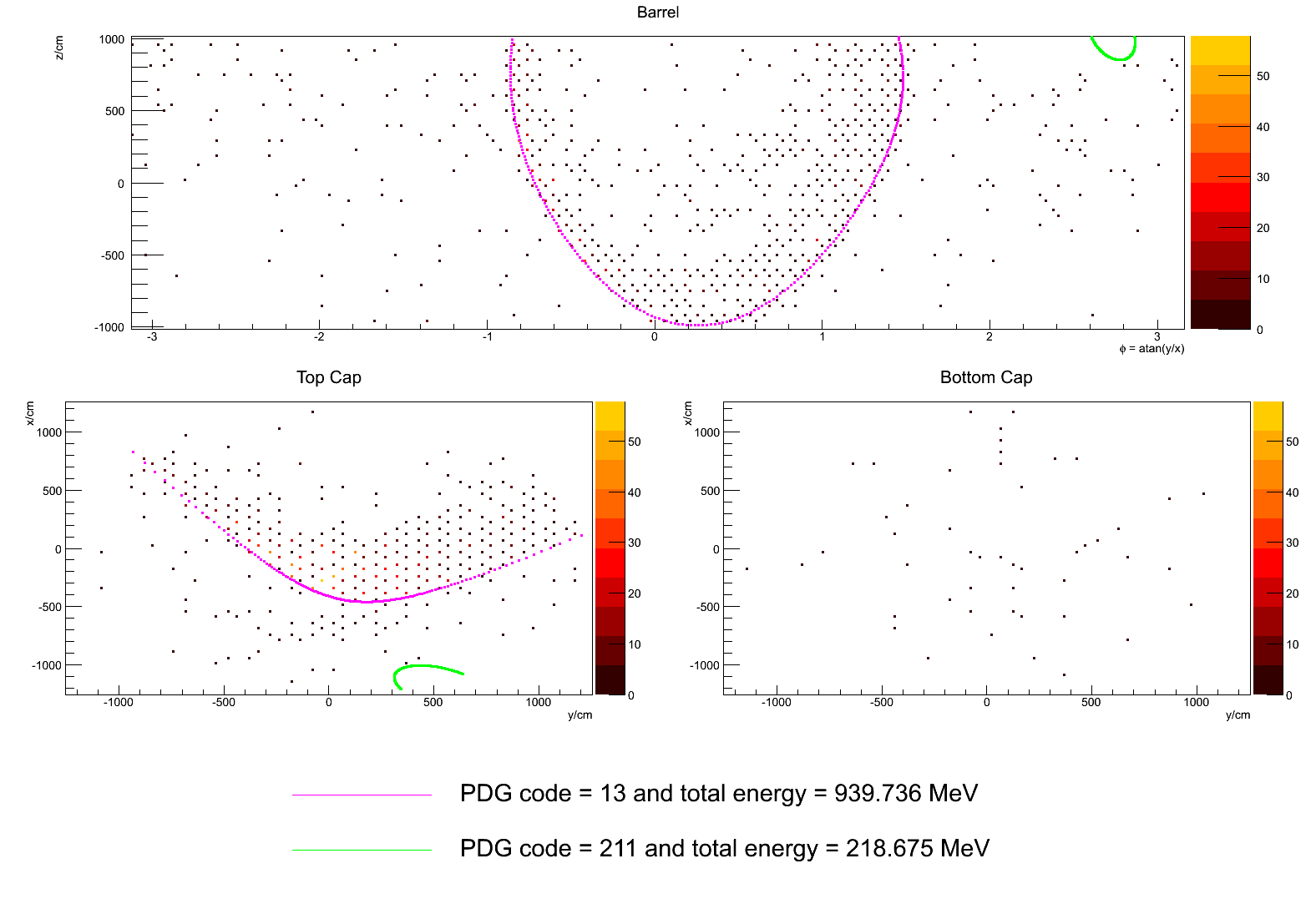}
\caption{Event display with rings drawn based on MC truth information for a
\numu CC interaction in a 10\kton CHIPS detector. Two PMT types are placed
in a diagonal pattern, with a total photocathode coverage of 10\%.}
\label{fig:eventdisplay}
\end{center}
\end{figure}

A Geant4\cite{Agostinelli:2002hh} simulation package has been written for CHIPS-10.  
Figure \ref{fig:eventdisplay} shows a sample event display from this simulation.  The
package is based on the simulation tool for the former LBNE water Cherenkov option,
with a number of major additions to suit CHIPS design studies.  In particular,
it allows multiple types of PMT to be placed in the same detector, with
photomultipliers arranged in any arbitrary repeating pattern on a square
lattice, instead of a simple square array.  This feature is illustrated in
Figure \ref{fig:layout_schematic}.

The detector is also divisible into regions, with the PMT layout of each region
specified separately.  This enables the testing of layouts with, for example, a
higher density of photomultipliers on the downstream region of the detector
relative to the beam.

\section{Summary} During the summer of 2014, the CHIPS collaboration
successfully constructed and deployed a small prototype water Cherenkov detector
at the bottom of a 60\m deep pit in the NuMI beam.  The detector has been taking
daily cosmic runs and measuring the rate of coincident hits between DOMs, and
has allowed tests of the water filtration system and environmental monitoring
vessels.  It will be retrieved in June 2015 to assess the durability of the
detector structure and geomembrane liner.  

Meanwhile, the design and simulation of a full 10\kton module are underway,
including features for complex PMT layouts to maximise the physics performance.

\end{document}